# Structure and Thermodynamics of Molecular Hydration via Grid Inhomogeneous Solvation Theory

Crystal Nguyen[1], Michael K. Gilson[1]*, and Tom Young[2]*

1. Skaggs School of Pharmacy and Pharmaceutical Sciences, U.C. San Diego, 9500 Gilman Drive, La Jolla, CA, 92093-0736. cnnguyen@ucsd.edu and mgilson@ucsd.edu

2. Dept. of Chemistry, Lehman College, CUNY, 250 Bedford Park Blvd. West, Bronx, NY 10468. simpleliquid@gmail.com

* To whom correspondence should be addressed

# Abstract

Changes in hydration are central to the phenomenon of biomolecular recognition, but it has been difficult to properly frame and answer questions about their precise thermodynamic role. We address this problem by introducing Grid Inhomogeneous Solvation Theory (GIST), which discretizes the equations of Inhomogeneous Solvation Theory on a 3D grid in a volume of interest. Here, the solvent volume is divided into small grid boxes and localized thermodynamic entropies, energies and free energies are defined for each grid box. Thermodynamic solvation quantities are defined in such a manner that summing the quantities over all the grid boxes yields the desired total quantity for the system. This approach smoothly accounts for the thermodynamics of not only highly occupied water sites but also partly occupied and water depleted regions of the solvent, without the need for *ad hoc* terms drawn from other theories. The GIST method has the further advantage of allowing a rigorous end-states analysis that, for example in the problem of molecular recognition, can account for not only the thermodynamics of displacing water from the surface but also for the thermodynamics of solvent reorganization around the bound complex. As a preliminary application, we present GIST calculations at the 1-body level for the host cucurbit[7]uril, a low molecular weight receptor molecule which represents a tractable model for biomolecular recognition. One of the most striking results is the observation of a toroidal region of water density, at the center of the host's nonpolar cavity, which is significantly disfavored entropically, and hence may contribute to the ability of this small receptor to bind guest molecules with unusually high affinities.

## Introduction

The treatment of water in structure-based drug-design poses practical challenges and conceptual conundrums[1-6]. These arise most prominently when the binding pocket of a targeted protein contains one or more crystallographic water molecules, as is frequently the case. In this setting, it is rarely clear on structural grounds whether the affinity of a proposed small molecule ligand will be maximized by displacing the water, using it as a bridge between the ligand and the protein, or avoiding contact with it altogether. It may even be unclear how to rigorously pose such questions. Indeed, a crystallographic water is not really a water, but a water site with water molecules exchanging in and out on some time-scale. Furthermore, the consequences of extending the ligand into a water site depend not only on the nature of the site but also on the ligand group that comes to occupy it. On the other hand, although sites with relatively low water density[7] will not be seen crystallographically, they, too, may have implications for ligand optimization; and, as discussed below, some binding pockets favor more complex multi-water structures that can affect ligand binding affinities. Thus, although crystallographically identified waters are of great interest, they are only the tip of the iceberg, as it were, because the rest of the water in a binding pocket also is perturbed relative to bulk, in a manner determined by the shape of the local protein surface and its patterning of hydrophilic and hydrophobic groups.

The first-order effects of water on biomolecules are, arguably, consequences of its high dielectric constant and of the hydrophobic effect[8,9]. Water's high dielectric constant makes it a good solvent for polar and ionized groups, and thus helps account for such fundamental phenomena as the solubility of salts, the ionization of many acids and bases in water, and the tendency of polar residues to lie at protein surfaces. It also greatly weakens charge-charge interactions relative to what they would be in vacuum. The hydrophobic effect drives the association of nonpolar surfaces in water, and is typically interpreted as resulting from the entropic cost of water's tendency to form structures that maintain energetically favorable water-water hydrogen bonds at hydrophobic surfaces[10,11]. These central solvent effects are captured rather well by continuum solvation models: the Poisson-Boltzmann[12-16] and Generalized Born[17-23] models are widely used to model electrostatic solvation effects at modest computational cost, and the hydrophobic effect may be approximated through surface area[24-27] or more detailed[28-30] continuum models.

However, continuum models do not capture the consequences of the specific size and directional hydrogen-bonding of water molecules. For example, a patch of a protein's binding pocket might present hydrogen bonding groups to the solvent that a continuum model would treat as fully solvated but that cannot be simultaneously satisfied by hydrogen bonds with water due to their mutual proximity. Also, it is not yet clear how to use a continuum model to address the questions raised above regarding crystallographic waters. One can use Molecular dynamics (MD) and Monte Carlo (MC) simulations with explicit water in order to account for the consequences of the molecular, rather than continuum, nature of water. This capability is of value in a range of applications, but simulations alone do not provide much insight into the local properties of water in a binding pocket; for this, further analysis tools are needed. To date, the tools developed for this purpose have largely focused on discrete, highly occupied water sites. In one such approach, thermodynamic integration and related methods have been used to compute the free energy change of extracting an isolated water molecule from a binding pocket[31-35]. A drawback of this approach is that it is not clear how it can be applied to more complex settings, such as ones involving partly occupied water sites or sites where a removed water molecule will immediately be replaced by another from the bulk. This approach also appears not to provide much mechanistic insight into the reasons why some waters are more stable than others.

Recently, a formulation of liquid state statistical thermodynamics termed inhomogeneous solvation theory[36,37] (IST) has been combined with MD simulations to study the thermodynamics of crystallographic waters[38-40] and of highly occupied water clusters identified from the simulations themselves[7,41-43]. In this approach, an MD trajectory is used to evaluate the single-body entropy of the water occupying the site, as well as its mean energetic interactions with the rest of the system. These quantities are then used to assess the stability of the water in the selected sites. This method provides valuable insight into the role of specific water sites in molecular recognition. For example, application to the binding pocket of streptavidin revealed five high-occupancy water sites with energetics similar to that of bulk water, but with geometric restrictions that lead to greatly reduced orientational and translational entropy[41]. The large entropic gain when these waters are ejected into the bulk presumably helps account for the remarkably high affinity of streptavidin for its ligand, biotin[44]. Analyses of other binding pockets demonstrate another scenario, in which a water molecule in a highly occupied site cannot form its full complement of hydrogen bonds. It has been estimated that ejecting even a single such frustrated water molecule from a solute cavity into the bulk can yield a biologically significant contribution to the free energy (well over 1 kcal/mol)[42,7,45-48]. Such phenomena, which are not well handled by continuum models, clearly need to be accounted for in projects aimed at designing high affinity ligands for targeted proteins. To date, however, implementations of IST have been limited to the analysis of high-occupancy water sites. As a consequence, they have not been able to provide information on weakly occupied sites or to another interesting and potentially important class of binding pocket regions, ones in which the water density is low, rather than high, relative to bulk[45]. The inability of existing IST implementations to handle such "dry" regions has motivated a proposal to supplement the method with an *ad hoc* term whose theoretical connection to IST is unclear[7].

More broadly, due to their inability to treat a range of densities, along with conceptual problems associated with partial displacement of water from high density regions, existing implementations of IST have not provided results that could be directly related to or compared with standard methods of computing and interpreting solvation, such as continuum models or thermodynamic integration with explicit solvent. The lack of connection to solvation free energies has made the results of IST difficult to interpret or assess in any detail.

In summary, there is still a need for a clear conceptual framework, tightly coupled with practical tools, for thinking about, modeling, and taking advantage of the structure and thermodynamics of water in protein binding pockets.

Here, we introduce a direct implementation of inhomogeneous solvation theory (IST) on a 3D grid around the solute of interest. This novel approach yields many advantages over existing implementations. For one thing, it integrates over the entire binding pocket and hence automatically and smoothly accounts not only for highly occupied water sites, but also for partly occupied and water-depleted regions, without any requirement for *ad hoc* terms drawn from other theories. It also for the first time enables a formally rigorous thermodynamic end-states analysis based on IST. This can be used to estimate solvation free energies of small molecules; it also can be used in a "before and after" mode to estimate changes in solvation energy and entropy when, for example, a ligand extends into a new region of a binding pocket. Such grid IST (GIST) solvation free energy calculations for the first time enable detailed characterization of IST via direct comparisons with experimental data and with other solvation models.

The molecule cucurbit[7]uril (CB7)[49] is a small, synthetic receptor which binds guest molecules in water with extraordinarily high affinities normally associated only with much larger biomolecules[50,51]. It is therefore useful as a simple, yet informative model for biomolecular recognition. One reason for its high affinities is the strong preorganization of the host and its guests, combined with their high degree of chemical complementarity[52]. However, it is also possible that something special about the structure and thermodynamics of the water in and around this unique host molecule helps it to achieve such high binding affinities. It is of particular interest to study the water within the rounded, hydrophobic central cavity of this pumpkin-shaped[53] receptor, so we have taken this as the initial application of the present IST method.

## Methods

The GIST method uses data from explicit solvent simulations, here MD, to evaluate and analyze leading terms of inhomogeneous solvation theory (IST)[36,37]. This is done by discretizing the analytic expressions of IST onto a three-dimensional (3D) grid that is fixed in the reference frame of the solute and extends several solvation layers into solution around the solute. The following subsections review the required theory, describe the approach taken to discretization, and detail the simulations and their analysis.

**Inhomogeneous Solvation Theory**

Like other liquid theories, IST relies on a transformation of integrals over molecular coordinates to integrals over distribution functions, leading to expressions for thermodynamic quantities that are expressed in terms of one-body, two-body, etc., correlation functions[36,54-57]. We focus on the lower-order terms because they are the most computationally tractable yet are expected to capture much of the physics. The following subsections briefly review IST in order to define notation and provide a basis for the discretization methodology.

*Solvation Entropy*

Following Lazaridis et al.[37,58,59], one may approximate the solvation entropy, $\Delta S_{solv}$, of a solute in a given conformation as:

$$\Delta S_{solv} = \Delta S_{sw} + \Delta S_{ww}$$

where $\Delta S_{sw}$ accounts for solute-water (sw) correlations and $\Delta S_{ww}$ for water-water (ww) correlations. Further limiting attention to the solute-water term, we write

$$\Delta S_{solv} \approx \Delta S_{sw} \equiv -k_B \frac{\rho^o}{8\pi^2} \int g_{sw}(\mathbf{r},\omega) \ln g_{sw}(\mathbf{r},\omega) d\mathbf{r} d\omega$$

where the approximation reflects the single body truncation of the entropy expansion, so that it accounts for only solute-water correlations; $k_B$ is Boltzmann's constant; $\rho^o$ is the number density (mean number of waters per unit volume) of bulk solvent; $g_{sw}(\mathbf{r},\omega)$ is the solute-water pair-correlation function in the solute frame of reference, such that $\rho^o g_{sw}(\mathbf{r},\omega)$ is the number density of water at location and orientation $(\mathbf{r},\omega)$, where $\mathbf{r}$ may be defined as the location of the oxygen atom and $\omega$ gives the orientation of the water in Euler angles; and the factor of $1/8\pi^2$ normalizes the orientational ($\omega$) integrals. The integrand is similar in form to the familiar $-\rho \ln \rho$ of the Gibbs/Shannon entropy. However, because the correlation function $g$ is normalized to the bulk density, it goes to unity for unperturbed (bulk) solvent, causing the entropy to go to zero in this case. This is why the expression yields solvation entropy, rather than total entropy. By the same token, the solute-solvent correlation function $g_{sw}$ approaches unity with increasing distance from the solute. As a consequence, the integrand decays to zero with distance from the solute, so that $\Delta S_{solv}$ may be approximated by a local integral around the solute.

The solute-water term, $\Delta S_{sw}$, is broken into intuitively meaningful and computationally tractable translation and orientational terms[60] by rewriting $g_{sw}(\mathbf{r},\omega)$ as the product of a translational distribution function and an orientational one conditioned on the position, $\mathbf{r}$:
$g_{sw}(\mathbf{r},\omega) = g_{sw}(\omega|\mathbf{r}) g_{sw}(\mathbf{r})$ so that:

$$\begin{aligned}
\Delta S_{sw} &= \Delta S_{sw}^{trans} + \Delta S_{sw}^{orient} \\
\Delta S_{sw}^{trans} &\equiv -k_B \rho^o \int g_{sw}(\mathbf{r}) \ln g_{sw}(\mathbf{r}) d\mathbf{r} \\
\Delta S_{sw}^{orient} &\equiv \rho^o \int g_{sw}(\mathbf{r}) S^\omega(\mathbf{r}) d\mathbf{r} \\
S^\omega(\mathbf{r}) &\equiv \frac{-k_B}{8\pi^2} \int g_{sw}(\omega|\mathbf{r}) \ln[g_{sw}(\omega|\mathbf{r})] d\omega
\end{aligned}$$

Here $S^\omega(\mathbf{r})$ is a localized orientational entropy relative to that of bulk solvent, and $\Delta S_{sw}^{orient}$ is the spatial integral of this quantity, weighted by the local water density.

To head off possible confusion, note that $\Delta S_{sw}$ is single-body in the sense that it involves just one set of water coordinates. These coordinates comprise translational and orientational degrees of freedom, $(\mathbf{r},\omega)$, which in turn comprise three Cartesian coordinates and three Euler angles, respectively. Thus, although the expression for $\Delta S_{sw}$ is single-body, it accounts for correlations among six coordinates and in this sense is higher than first-order. Also, as noted above, the above entropies are referenced to a homogenous, unperturbed fluid because the translational and orientational correlation functions $g(\mathbf{r})$ and $g(\omega|\mathbf{r})$ are unity for the unperturbed solvent.

*Solvation Energy*

For a pairwise additive energy functions, such as the SPC/E[61] or TIP4P[62] water models, the full IST expansion of the solvation energy[63] conveniently terminates with the two-body term:

$$\Delta E_{solv} = \Delta E_{sw} + \Delta E_{ww}$$

$$\Delta E_{sw} \equiv \int \Delta E_{sw}^{loc}(\mathbf{r}) d\mathbf{r}$$

$$\Delta E_{sw}^{loc}(\mathbf{r}) \equiv \frac{\rho^o}{8\pi^2} \int g_{sw}(\mathbf{r},\omega) U_{sw}(\mathbf{r},\omega) \, d\omega$$

$$\Delta E_{ww} \equiv \rho^o \int g_{sw}(\mathbf{r}) \Delta E_{ww}^{loc}(\mathbf{r}) d\mathbf{r}$$

$$\Delta E_{ww}^{loc}(\mathbf{r}) \equiv \left(\frac{1}{8\pi^2}\right)^2 \frac{\rho^o}{2} \int g_{sw}(\mathbf{r},\omega)\left[ g_{sw}(\mathbf{r}',\omega') g_{ww}^s(\mathbf{r},\omega,\mathbf{r}',\omega') - g_{ww}^o(\mathbf{r},\omega,\mathbf{r}',\omega') \right] U_{ww}(\mathbf{r},\omega,\mathbf{r}',\omega') \, d\omega d\mathbf{r}' d\omega'$$

The two terms on the first line represent, respectively, the mean water-solute interaction energy and the mean change in water-water interaction energy due to addition of the solute to the water; the *loc* superscript is used to indicate an interaction energy associated with the water density at location *r*; $U_{sw}(\mathbf{r},\omega)$ is the solute-water interaction potential; and $U_{ww}(\mathbf{r},\mathbf{r}',\omega,\omega')$ is the water-water potential (which could be rewritten in terms of the relative water-water coordinates). The integrands decay with distance from the solute, so the solvation energies can again be approximated by integrals local to the solute. Note that the water-water interaction term introduces the risk of double-counting interactions, so care must be taken in defining and calculating it.

**Discretization of Inhomogeneous Solvation Theory**

Discretization of the IST equations provided above allows a grid-based numerical implementation of the theory, in which each integral is approximated as a sum over local quantities, and the local quantities are computed from the stored frames (snapshots) of an MD simulation. We limit attention here to the low-order terms discussed above. Although higher order terms can be similarly discretized in a straightforward manner, they require considerably longer simulations to reach adequate numerical convergence.

*Translational entropy*

We consider a volume that includes the solute and a solvent region of interest around it to be discretized into grid boxes $k$ of volume $V_k$ centered at locations $\mathbf{r}_k$, and then write the translational entropy as a sum over the local translational entropies for each grid box $k$ ($\Delta S_{sw}^{trans}(\mathbf{r}_k)$):

$$\Delta S_{sw}^{trans} = \sum_k \Delta S_{sw}^{trans}(\mathbf{r}_k)$$

$$\Delta S_{sw}^{trans}(\mathbf{r}_k) \equiv -k_B \rho^o \int_k g(\mathbf{r}) \ln g(\mathbf{r}) d\mathbf{r}.$$

$$\approx -k_B \rho^o g(\mathbf{r}_k) \ln g(\mathbf{r}_k)$$

where the integral is over the volume of grid box $k$, $V_k$ and the approximation consists of treating $g(\mathbf{r})$ as uniform within each grid box, with a value estimated from a finite number of simulation frames as:

$$g(\mathbf{r}_k) = \frac{1}{\rho^o N_f} \sum_{i=1}^{N_f} \frac{n_{i,k}}{V_k}$$

Here, $i \in 1 \ldots N_f$ indexes the simulation frames, and $n_{i,k}$ is the number of waters within box $k$ for frame $i$. A water molecule is considered to lie in box $k$ if the center of its oxygen atom is in the box.

*Orientational Entropy*

Similarly, we write the orientational entropy as a sum over grid boxes, typically but not necessarily the same ones as those used for the translational entropy, and we treat both $g(\mathbf{r})$ and the localized orientational entropy, $S^\omega(\mathbf{r}_k)$, of waters as uniform across box $k$:

$$S^\omega(\mathbf{r}_k) = \frac{-k_B}{V_k 8\pi^2} \int_k \int_\omega g(\omega|\mathbf{r}) \ln g(\omega|\mathbf{r}) d\omega d\mathbf{r}$$

.

This approximates $g(\mathbf{r})$ as uniform within box $k$, so that the local orientational entropy within it is uniformly weighted in the integral over $\mathbf{r}$. Given adequate sampling, the approximation becomes exact in the limit $V_k \to 0$ and is clearly more accurate for smaller grid boxes.

This allows us to write the density weighted orientational entropy as:

$$\Delta S_{sw}^{orient}(\mathbf{r}_k) = \rho^o \int_k g(\mathbf{r}) S^\omega(\mathbf{r}_k) d\mathbf{r}$$

.

And the total overall entropy for the system as:

$$\Delta S_{sw}^{orient} \approx \sum_k \Delta S_{sw}^{orient}(\mathbf{r}_k) = \sum_k n_{k,av} S_k^\omega$$

$$n_{k,av} \equiv \sum_{i=1}^{N_f} \frac{n_{i,k}}{N_f}$$

Here $n_{k,av}$ is the mean number of waters found in box $k$ across the $N_f$ simulation frames.

The value of $S^{\omega}(\mathbf{r}_k)$ can be computed from simulation data by various means, such as histogram and nearest-neighbor[64-66] methods. We use histograms in this initial implementation. This is done by computing Euler angles $(\theta, \phi, \psi)$, in the solute frame of reference, for each water found in the box across all frames, and binning these triplets in uniformly by $\cos\theta, \phi, \psi$. The count in each histogram bin was normalized by the total number of waters in the histogram and divided by the orientational bin volume $\Delta(\cos\theta)\,\Delta\phi\,\Delta\psi$ to yield a probability density which goes to $(8\pi^2)^{-1}$ in all bins for a uniform orientational distribution.

*Energy*

The solute-water interaction energy is readily, and exactly, decomposed into a sum over grid boxes, *k*, as follows:

$$\Delta E_{sw} = \sum_k \Delta E_{sw}(\mathbf{r}_k)$$
$$\Delta E_{sw}(\mathbf{r}_k) \equiv \int_k \Delta E_{sw}^{loc}(\mathbf{r}) d\mathbf{r}$$

Similarly, the water-water energy for the system is decomposed as

$$\Delta E_{ww} = \sum_k \Delta E_{ww}(\mathbf{r}_k)$$
$$\Delta E_{ww}(\mathbf{r}_k) \equiv \rho^o \int_k g_{sw}(\mathbf{r}) \Delta E_{ww}^{loc}(\mathbf{r}) d\mathbf{r}$$

In practice, the value of $\Delta E_{sw}(\mathbf{r}_k)$ is computed as the solute interaction energy, according to the simulation force field, of all water molecules in box $k$, averaged over the simulation frames. Similarly, $\Delta E_{ww}(\mathbf{r}_k)$ is computed as the interaction energy of all waters in box $k$ with all other waters in the system, averaged over all frames.

*Free Energy*

With the localized orientational entropy ($\Delta S_{sw,k}^{orient}$), translational entropy ($\Delta S_{sw,k}^{trans}$), and the localized energy ($\Delta E_{sw}(\mathbf{r}_k)$), we can define a local population weighted local free energy of solvation ($\Delta G_{sw}(\mathbf{r}_k)$) in an arbitrarily sized volume $V_k$ as:

$$\Delta G_{sw}(\mathbf{r}_k) = \Delta E_{sw}(\mathbf{r}_k) - T \Delta S_{sw}^{total}(\mathbf{r}_k)$$

where $T$ is the temperature and the total localized entropy, $\Delta S_{sw}^{total}(\mathbf{r}_k)$, is just the sum of the local orientational and translational entropies:

$$\Delta S_{sw}^{total}(\mathbf{r}_k) = \Delta S_{sw}^{trans}(\mathbf{r}_k) + \Delta S_{sw}^{orient}(\mathbf{r}_k).$$

The Total Free Energy for the system can then be written as:

$$\Delta G_{sw} = \sum_k \Delta G_{sw}(\mathbf{r}_k).$$

**Computational Details**

Force-field parameters of the synthetic host molecule cucurbit[7]uril (CB7) (Figure 1) were generated as follows. Partial charges were computed with the program RESP[67], part of AMBER 11[68], based on electrostatic structure calculations at the 6-31G* level with the program Gaussian 2003. The remaining parameters were assigned from the GAFF[69] force field, with the AmberTools program xleap. This circular host molecule, which is about 13Å in diameter, was then computationally immersed in a 40Å x 43Å x 43Å box of pre-equilibrated TIP4P water molecules, using the program Leap. The resulting system consists of 126 solute atoms and 1699 water molecules. All simulations were carried out with a graphical processor unit-enabled version of the Amber program PMEMD[68] using Langevin dynamics[70] with a collision frequency of 2 ps$^{-1}$, periodic boundary conditions, a nonbonded cutoff distance of 9.0 Å coupled with Particle-Mesh Ewald long-ranged electrostatics, a time step of 2fs, and SHAKE[71] for covalent bonds to hydrogen atoms. Each Langevin simulation used a different random number seed. Center-of-mass translation of the host was removed every 1000 steps to keep the system centered. For the equilibration and production simulations, constant 1 atm pressure was maintained with isotropic positional scaling and a relaxation time of 0.5 ps to ensure that the system density remained appropriate. Trajectory frames for analysis were saved at 0.5 ps intervals.

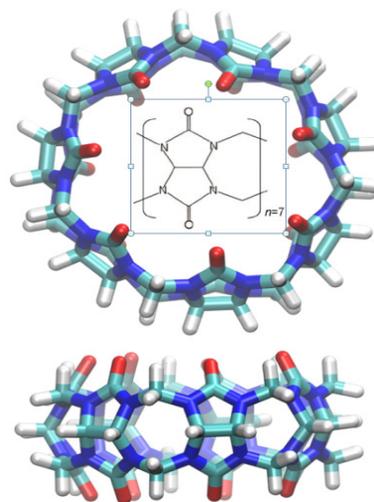

Figure 1. Top and side views of the cucurbit[7]uril host molecule. Top view also shows the repeating chemical unit.

The initial system was relaxed with 1500 cycles of steepest descent followed by 500 cycles of conjugate gradient energy-minimization. The relaxed system was then heated gradually to 300K in steps of 50K, each lasting 20ps. The system was then equilibrated for 5ns at 300K, and a 40 ns production run was then carried out. This procedure, starting from the minimized configuration, was repeated 10 times with different random number seeds to generate 10x40=400ns of simulation time.

The GIST methodology was implemented on a 3D rectangular grid centered on the host, as detailed below. The coordinates of each trajectory frame were registered to the grid by repositioning the host's center of mass to the origin and removing rotations of the host; water molecules and their images were also relocated accordingly. The grid spacing was 0.5 Å along each axis, for a grid box volume of 0.125 Å$^3$. This spacing was found to provide a detailed representation and good convergence of local water density. Local thermodynamic quantities associated with the grid boxes were written to disk in the form of Data Explorer (dx) files, to enable visualization, in the context of the host molecule, with the program VMD[72]. The orientational histograms divided the $\cos\theta$, $\phi$ and $\psi$ axes into 11 steps each, for a total of 1,331 orientational bins.

This initial application of the GIST approach focuses on the translational and orientational entropy of the water in and around the host molecule CB7, along with the water-host interaction energy. The water-water interaction energies, which are more difficult to define in a local manner, will be addressed in subsequent work.

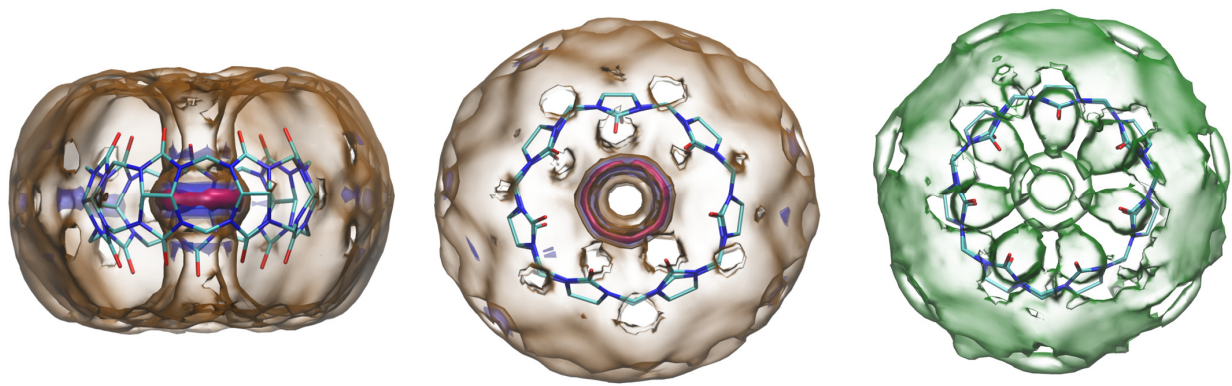

Figure 2. Side and top views of CB7 with water oxygen density contours at 5x (red), 3x (blue) and 1.5x (glassy), followed by a contour of water hydrogen density highlighting the localization of water hydrogens in the vicinity of the hosts' carbonyl oxygens.

## Results

Figure 2 shows the solvent density around CB7 with contours at factors of 5 (red), 3 (blue) and 1.5 (glassy) times bulk density. Perhaps the most striking features of the density distribution are a sharply defined central toroid of particularly high density and two less-crisp, lower-density toroids above and below the central toroid. The first peak of the oscillatory radial distribution function around the host lies between the two layers of the g=1.5 contour discernable here. There are no obvious peaks in solvent density associated with the 14 carbonyl oxygens, whereas water hydrogens clearly cluster over the carbonyls (Figure 2).

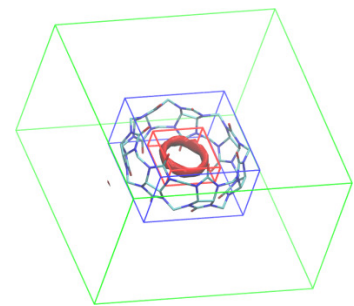

Figure 3: CB7 host with water density torus and boxes indicating the three regions of interest: Grid (green), Cavity (blue) and Torus (red).

We dissected the solvation structure and thermodynamics by accumulating GIST results across three regions shown in Figure 3: the entire 3D GIST grid, a subgrid containing the cavity of the host, and a smaller subgrid that includes mainly the water density of the central torus. Table 1 lists the mean number of water molecules in each region, followed by the changes in 1-body translational and orientational entropy relative to bulk solvent, and the mean solvent-host interaction energy.

|  | $n_{av}$ | $-T\Delta S_{trans}$ | $-T\Delta S_{orient}$ | $\Delta E_{sw}$ | Sum |
|---|---|---|---|---|---|
| **Grid** | 208 | 13.4 | 55 | -44.6 | 23.8 |
| **Cavity** | 6.7 | 3.3 | 5.7 | -7.9 | 1.1 |
| **Torus** | 3.5 | 2.4 | 1.2 | -2.7 | 0.9 |

Table 1. Mean number of water molecules $n_{av}$, and thermodynamic quantities from GIST (see text) for the entire 3D Grid, the Cavity subgrid, and the Torus subgrid, shown in Figure 3. Entropies and energies are based upon 400 ns and 200ns of simulation time, respectively. The rightmost column provides the sum of the thermodynamic quantities. Units of -TS and E are kcal/mol.

The convergence of the entropic quantities in Table 1 as a function of simulation time is displayed in Figure 4. The translational entropies, which derive from the water density distributions are well converged, as are the orientational entropies for the Cavity and Torus regions. However, the orientational entropy across the entire grid is still change at a substantial rate after 400 ns of simulation time. The water-solute interaction energies ($\Delta E_{sw}$) converge rapidly, however, becoming stable to within 0.1 kcal/mol of their final values are well-converged for all three regions.

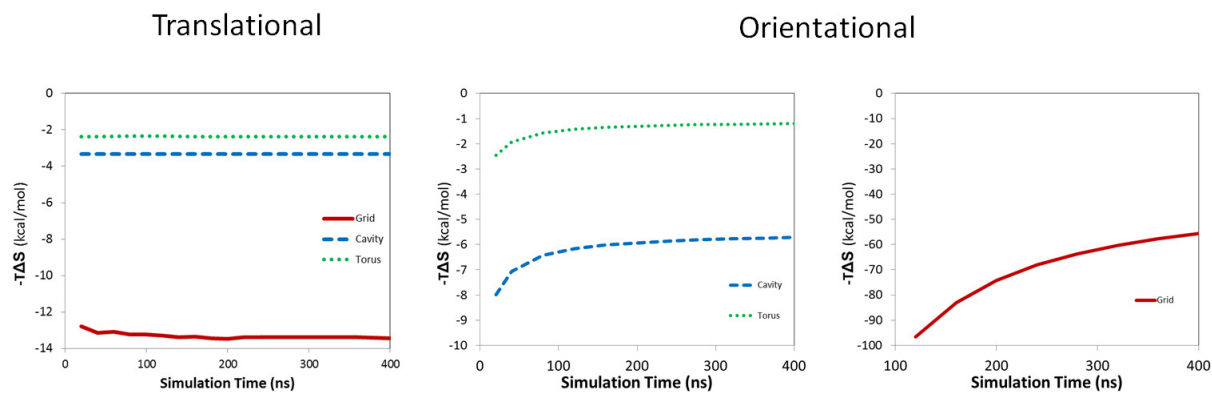

Figure 4. Convergence of entropic quantities in Table 1 as a function of simulation time. The orientational results are split across two graphs so the total grid orientational entropy can be presented on its own scale.

The richness of the information about solvation that is provided by the GIST approach is highlighted by 3D contours of the local quantities $\Delta E_{sw}(\mathbf{r}_k)$ and $-T\Delta S_{sw}^{orient}(\mathbf{r}_k)$, shown in Figure 6, and this system will be further analyzed in a subsequent publication.

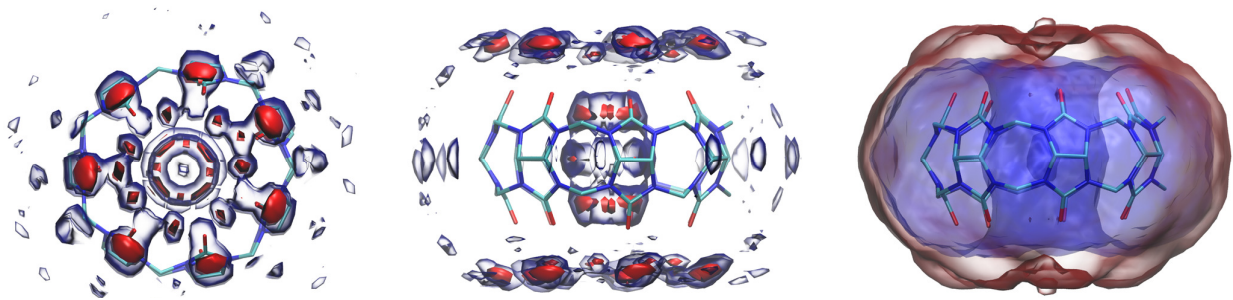

Figure 5. Contours of $\Delta E_{sw}(\mathbf{r}_k)$ (top view on left and side view at center) show multiple sites of stabilizing water interactions with the carbonyl oxygens of the host CB7. Contours of $-T\Delta S_{sw}^{orient}(\mathbf{r}_k)$ (right) show increased ordering closer to the solute, and especially in the low-density core at the very center of the cavity.

## Discussion

We have described the GIST framework for modeling solvation structure and thermodynamics, and demonstrated initial application to the low molecular weight receptor CB7. This molecule was chosen because it is conveniently simple, yet captures many of the essential features of biomolecular recognition. The calculations reveal remarkable intricacy in the solvation features of even this small, symmetric solute. One of the most interesting findings is a sharply localized torus of high solvent density at the center of the host's rather nonpolar cavity. The mere localization of this water here is estimated to generate 2.4 kcal/mol worth of entropy relative to bulk water, while orientational ordering is estimated to generate another 1.2 kcal/mol entropic penalty. Thus, the estimated entropic cost of forming this toroidal water feature is about 3.6 kcal/mol, at the 1-body level. These waters are stabilized by about -2.7 kcal/mol of energetic interaction with the host, and undoubtedly also by water-water interactions not reported here. However, the water-water interactions seem unlikely to be as strong as those of water molecules in the bulk, as the torus waters are relatively isolated. One may use these data to consider the consequences of this water feature for the binding of a guest molecule in the cavity of CB7, and hence expulsion of the water torus. A guest molecule will form its own stabilizing interactions with the interior of the cavity, balancing the -2.7 kcal/mol interaction energy of the water, and the expelled waters should gain on the order of -3.6 kcal/mol worth of entropy. This favorable entropy change will contribute to the affinity of the guest molecule and may be one of the factors contributing to the ability of CB7 to bind guests with extraordinarily high affinities normally associated only with biomolecules. We anticipate future applications of the GIST approach to not only host-guest chemistry but also protein-ligand binding and, in particular, computer-aided drug design, especially as this novel method offers important advantages. For one thing, it yields an integral over the entire region of interest, such as a protein binding pocket, and hence automatically and smoothly accounts not only for highly occupied water sites, but also for partly occupied and water-depleted regions, without any requirement for *ad hoc* terms drawn from other theories. Existing approaches to this problem may not account for all the water molecules in a system and may encounter conceptual problems related to how many water molecules are accounted for in a hydration site[73]. Although not explored here, GIST also provides a mechanism by which molar free energies of solvation can be estimated by summing well-defined quantities over the whole grid. This offers the possibility of using IST to estimate solvation free energies, thereby connecting directly with the literature on solvation theory and models. Finally, as sketched below for the present CB7 application, the localized chemical potential of solvation provided by GIST can be used to identify solvation hotspots -- regions of solvation that are thermodynamically unfavorable -- from which it would be thermodynamically favorable to eject water. We anticipate that this type of analysis will be particularly useful to characterize solvent properties in protein binding sites and thus guide the design of high affinity ligands as drug candidates. Another unique feature of the approach is that it for the first time enables the use of "before" and "after" IST calculations to estimate changes in solvation energy and entropy when, for example, a ligand extends into a new region of a binding pocket. Such an end-states analysis consists of an initial MD calculation of a system in state **A** that is used to compute a "before" value of the total solvation free energy via IST, $\Delta G_{solv,A} = \Delta E_{solv,A} - T\Delta S_{solv,A}$. A second MD of system in state **B** is then used to compute an "after" value of the total solvation free energy $\Delta G_{solv,B} = \Delta E_{solv,B} - T\Delta S_{solv,B}$. The IST estimate of the change in solvation free energy for the two states is simply $\Delta\Delta G_{IST} = \Delta G_{solv,B} - \Delta G_{solv,A}$. This type of study could be particularly useful in the field of drug design where it can be applied to calculate the difference in free energy upon modification of a lead compound, where the protein complex with unmodified ligand defines state A, and the complex with modified ligand defines state B. One might similarly, though more ambitiously, estimate the free energy of expelling solvent from a binding site, where the ligand-protein complex is be state A and the apoprotein is state B.

This initial study opens a number of directions for future work. For the CB7 system, it will be of great interest to study the properties of other waters within the host cavity, the structure and thermodynamics of water around the carbonyl oxygens at both portals of the host, and the magnitude and role of water-water interactions throughout the system. For the GIST approach, one specific next step is to mitigate the convergence difficulties manifest here for the orientational entropy by substituting the nearest-neighbor method[64-66,74] for histograms. More broadly, the method needs to be put to the test in a variety of applications and molecular systems.